\documentclass[aps,prb,twocolumn,10pt,groupedaddress,showpacs,showkeys]{revtex4-1}
\usepackage[T1]{fontenc}
\usepackage{lmodern}
\usepackage{graphicx}

\begin{document}


\title{Ferromagnetism in YbCu$_2$Si$_2$ at high pressure}


\author{A. Fernandez-Pa\~nella, D. Braithwaite, B. Salce, G. Lapertot and J. Flouquet}
\affiliation{SPSMS, UMR-E CEA / UJF-Grenoble 1, INAC, 38054 Grenoble, France}
\email [] {Corresponding author: daniel.braithwaite@cea.fr}



\begin{abstract}
We demonstrate from detailed ac susceptibility and calorimetry studies under hydrostatic pressure that YbCu$_2$Si$_2$ probably orders ferromagnetically at high pressure. The (\itshape{p,H,T}\normalfont) phase diagram shows that the transition temperature increases with pressure but also with an applied magnetic field. We suggest that many ytterbium systems may show a trend towards ferromagnetism and we discuss the possible reasons for this. We also examine the implications, including the potential of YbCu$_2$Si$_2$ and other Yb compounds for further studies of the rich physical properties that may occur near a ferromagnetic critical point.
\end{abstract}


\maketitle
\section{Introduction}
A key point for the understanding of quantum phenomena at a magnetic point is the comaprison between cerium and ytterbium systems. Both Yb and Ce can assume a magnetic (Ce$^{3+}$,4f$^1$/Yb$^{3+}$,4f$^{13}$) configuration or fluctuations between this and a nonmagnetic (Ce$^{3+}$,4f$^1$/Yb$^{3+}$,4f$^{13}$) state. The cleanest way to tune a system to its critical point is with high pressure. In both families pressure tends to decrease the occupancy of the 4f shell, leading to opposite effects on the magnetism. In cerium systems pressure tends to drive the system from magnetic order to a paramagnetic ground state whereas in ytterbium systems magnetic order tends to be induced by pressure. The free Yb$^{3+}$ ion (J=7/2) with 13 electrons in the 4f shell can be considered as the hole analogue of Ce$^{3+}$. In a real lattice, however, there are important differences, as summarized in a recent review\cite{Harima09}: the deeper localization of the 4f electrons, and the stronger spin-orbit coupling in Yb can lead to a different hierarchy of the significant energy scales (Kondo temperature, Crystal field). Moreover, in Yb the valence state is able to change far more between 2 and 3, whereas in cerium systems the range is generally much more restricted close to 3. Yb systems are therefore rather easily in an intermediate valence state with fluctuations between the integer values, and this can be the case even when long range magnetic order is present.\cite{Harima09} \\However a more basic element in order to compare the physics at a critical point of any ytterbium system with its cerium counterpart is the nature of the order that occurs. In the cerium systems where quantum criticality is induced, the magnetically ordered phase is established at ambient pressure. It can therefore be fully characterized by neutron scattering and magnetization measurements and in most known cases is antiferromagnetic (AFM). On the other hand in ytterbium where a critical point is induced with pressure, the magnetic order will appear under pressure, often too high for neutron scattering, and no direct determination of the magnetic structure is possible. There is some evidence that in several ytterbium compounds ferromagnetic (FM) correlations might dominate. YbInNi$_4$ orders ferromagnetically \cite{Sarrao98} or at least with a strong FM component.\cite{Willers} In YbInCu$_4$ magnetic order is induced at rather low pressure (3 GPa) which has allowed magnetization measurements showing a FM state.\cite{Mito2007} In other systems evidence is more indirect: FM fluctuations have been shown in YbRh$_2$Si$_2$\cite{Gegenwart2005} from transport measurements it was also suggested that the order that appears at high pressure (8 GPa) in YbRh$_2$Si$_2$ and YbIr$_2$Si$_2$ might be FM.\cite{Yuan2006}
Magnetoresistance at high pressures also point to possible FM correlations and order in YbNi$_2$Ge$_2$.\cite{Knebel2001} YbCu$_2$Si$_2$ is one of the clearest cases where magnetic order can be induced with pressure, this having been confirmed by resistivity,\cite{Yadri98} calorimetry,\cite{Colombier2009} and Mössbauer studies.\cite{Winkelman99} However none of these techniques can give direct information on the type of order, nor distinguish between AFM and FM order. We show here ac susceptibility measurements under high pressure which show direct evidence for FM order in this system.

\section{Experimental details}
High-quality single crystals were grown by an indium flux method (using MgO crucibles) as described in detail elsewhere.\cite{Colombier2009} The crystals were characterized by resistivity and the residual resistivity ratio (RRR$_{0.7-300K}$= 220) attests to the excellent quality with a corresponding residual resistivity of 0.3 ${\mu\Omega}$. Measurements under pressure were carried out using diamond-anvil cells. As pressure transmitting medium we used argon which has excellent hydrostatic conditions up to 10 GPa. Measurements were performed between 1.5 K to 20 K and from nearly ambient pressure to 13 GPa in a He cryostat with the \itshape in situ \normalfont pressure-tuning device.\cite{Salce2000} Pressure was determined from the ruby fluorescence by placing some ruby chips in the pressure chambers. The ac susceptibility technique we used is an adaptation of the technique developed in Cambridge.\cite{Alireza2003} We directly placed the sample inside the pickup coil (350 $\mu$m of external diameter, 10 turns with a 12 $\mu$m insulated Cu wire). An ac field of 0.1 mT was generated by the primary coil outside the pressure chamber at a frequency of 721 Hz. The magnetic field was applied parallel to the easy \itshape c \normalfont axis of  magnetization.\cite{Muz98,dung2009} Specific heat was measured by an ac calorimetry technique detailed elsewhere.\cite{Demuer2000} The sample was heated by a laser modulated by a mechanical chopper at about 722 Hz which was found, from the signal-frequency dependence, high enough to thermally decouple the sample from its environment. The temperature oscillations which are inversely proportional to the specific heat were obtained via the voltage measured from the Au/AuFe thermocouple soldered directly on the sample and its thermoelectric power. A small constant magnetic field up to 0.45 T could be superposed using a superconducting magnet.

\section{Results and disccussion}
\subsection{Ac susceptibility and ac calorimetry under pressure}
Figure 1(a) shows the low-temperature dependence of the ac susceptibility at selected pressures. For pressures above 6.5 GPa we find the onset of an increase of the ac susceptibility ($\chi'$) for T < 4 K, which becomes a clear peak for p > 8 GPa corresponding to the magnetic transition which we followed up to nearly 13 GPa. The broadening of the peak and the decreasing intensity of the signal with increasing pressures could be caused by homogeneity effects inside the pressure chamber. Unlike with a dc magnetization measurement, distinguishing between a FM and an AFM transition with ac susceptibility is not as trivial due to irreversibility effects in the FM state $\chi'$ can decrease with decreasing temperature below the Curie temperature (T$_{Curie}$) and frequently shows a peak as found here. Indeed a similar effect was seen in $\chi'$ in YbInCu$_4$.\cite{Mito2003}  In order to determine the nature of the magnetic order we have therefore made an estimate of the absolute value of the susceptibility. This was done using two methods. 
\begin{figure}[!htbp]
\includegraphics[height=5.5cm]{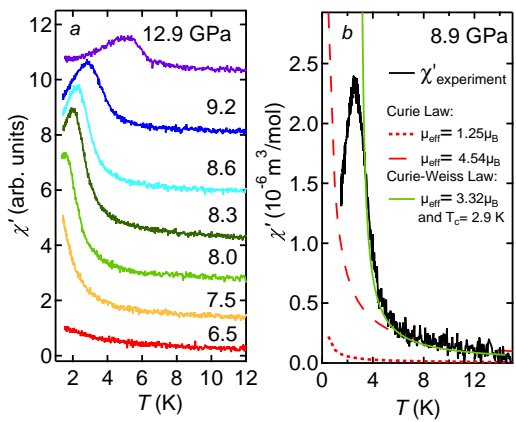}
\caption{(Color online) (a) ac susceptibility versus temperature for different pressures below and above the critical pressure. (b) Susceptibility at 8.9 GPa compared with expected Curie dependences for: $\mu_{eff}$ = 4.54 $\mu_B$ (dashed  line) and  $\mu_{eff}$ = 1.25 $\mu_B$ (dotted line).}
\end{figure}
First we calculated the value using estimations of the sample and pick-up coil geometries and filling factor.\cite{Alireza2003} Second we compared the signal from the sample with the superconducting transition of a Pb sample with the same dimensions in the same set-up. The two methods gave similar results with less than 20 \%difference. 
In Figure 1(b) the susceptibility at 8.9 GPa is compared with two Curie laws (red lines) and a Curie-Weiss law (green line). The experimental $\chi'$ contains an unknown but roughly constant background, so the zero has been taken at the highest temperature measured (15 K). First we compare the experimental curve with the Curie law corresponding to the effective moment of $\mu_{eff}$ = 1.25 $\mu_B$ which was determined experimentally by Mössbauer spectroscopy at a similar pressure.\cite{Winkelman99} This value is similar to that found in other Yb systems with the tetragonal (ThCr$_2$Si$_2$) structure showing magnetic order and compatible with the moment expected for Yb$^{3+}$ taking into account the effects of the crystalline electric field (CEF) and Kondo screening.\cite{Bonville91,Bonville92,Hodges87} The experimental data show clearly a much larger magnetic signal. We also show the Curie law for an effective magnetic moment corresponding to the theoretical value for the Yb$^{3+}$ free ion, $\mu_{eff}$ = 4.54 $\mu_B$. Even in this case the experimental data still diverges much more quickly than the theoretical curve below 5 K. The agreement of the experimental data above 5 K is probably coincidental and can also be compatible with a smaller moment and a positive Curie-Weiss temperature. Finally, the green line shows the best fit for this data with a Curie-Weiss law (with T$_{CW}$ = 2.9 K and $\mu_{eff}$ = 3.52 $\mu_B$). This value for $\mu_{eff}$ is surprisingly large. This could be due to a combination of pressure and sample inhomogeneity, and the very strong slope of dT$_M$/dP, causing a spread of ordering temperatures within the sample. All these points strongly lead toward ferromagnetism. 
\begin{figure}[!htbp]
\includegraphics[height=5.5cm]{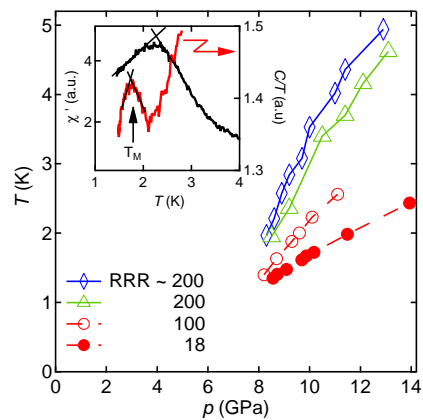}
\caption{(Color online) (\itshape {p,T}\normalfont) phase diagram: Diamonds and triangles represent T$_M$ obtained from ac susceptibility and ac calorimetry respectively in this study. Circles are the results from previous measurements (full circles Ref.8 and open circles Ref.9). Lines are guides for the eyes. The inset shows the criterion used to determine T$_M$. The improvement in sample quality (higher RRR) seems to lead to systematically higher ordering temperatures}
\end{figure}
Of course from these results we cannot exclude a more complex magnetic structure like canted antiferromagnetism as recently suggested in YbInNi4\cite{Willers} but we need at least a significant FM component. The pressure dependence of the magnetic temperature T$_M$ obtained by ac-susceptibility is summarized in Figure 2 and agrees well with the data found from calorimetry measurements on the same batch. Interestingly the ordering temperatures are found to be much higher than reports on previous batches of crystals, possibly reflecting the improved crystal quality. 

\subsection{Magnetic field effect on the ac susceptibility and ac calorimetry under pressure}
Further proof of FM order is found in the behavior when a small constant magnetic field is applied along the \itshape c\normalfont-axis. Susceptibility and specific heat curves for selected pressure around 8.5 GPa are shown in Figure 3 (a) and (b) respectively. In both cases, the amplitude of the transition decreases with the application of low magnetic fields but at the same time, T$_M$ is clearly shifted towards higher values for increasing fields. This trend can be better observed in the phase diagram (T$_M$, H) of Figure 3 (d) were we have plotted the transition temperatures versus the applied magnetic fields. 
\begin{figure}[!htbp]
\includegraphics[height=6.5cm]{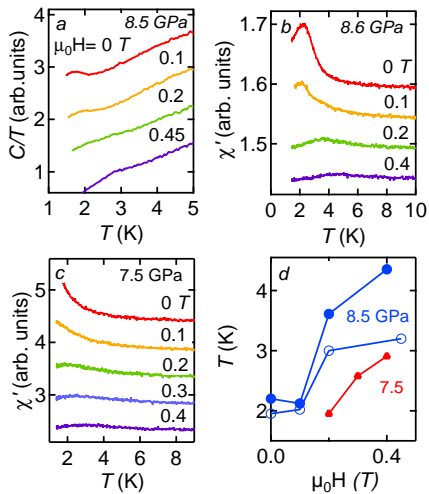}
\caption{(Color Online) Effect of magnetic field: (a) ac calorimetry curves at 8.5 GPa and magnetic fields up to 0.45 T. (b) Ac susceptibility vs temperature at 8.6 GPa and selected magnetic fields. (c) Susceptibility curves at 7.5 GPa under magnetic field. The transition is induced for fields greater than 0.2 T. (d) Magnetic dependence of T$_M$ near 8.5 GPa (blue curves) and 7.5 GPa (red curve) obtained from specific heat (open circles) and susceptibility (full circles and triangles) measurements.}
\end{figure}
This response, where T$_{Curie}$ increases with field, is consistent with the usual behaviour of a ferromagnetic system under magnetic field\cite{Willers}, and is contrary to the usual behavior of an antiferromagnet. Surprisingly, as shown in Figure 3(c), at 7.5 GPa where no transition is visible down to 1.5 K at zero field we found we could induce a metamagnetic transition by the application of dc fields above 0.2 T. The rapid decrease of the intensity in $\chi'$ is similar to the effect seen in YbInCu4,\cite{Mito2003} and probably mainly reflects the loss of any ac response in the FM state once a single domain has been created with the dc field, as well as the fact that under dc field the transition becomes a crossover as the magnetic field also breaks time-reversal symmetry from a partially to fully polarized state. 

A complete phase diagram of the (\itshape p, H, T\normalfont) phase diagram of T$_M$ is shown in Figure 4. We can observe that the trend of the increase of T$_M$ under magnetic field at 8.5 GPa is valid for all pressures we measured. Interestingly we find that T$_M$ does not significantly change for any pressure at very low magnetic fields (up to 0.1 T). The reason for this is not clear, but it could be that this is the field necessary to obtain a single domain and that below this value the internal field depends weakly on the applied field. 

\begin{figure}[!htbp]
\includegraphics[height=6cm]{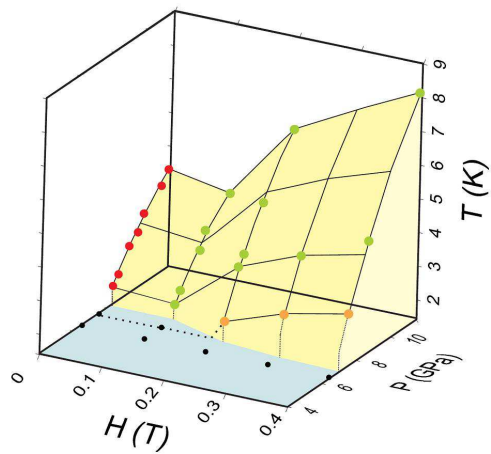}
\caption{(Color online) Complete Phase diagram (\itshape {p, H, T$_M$}\normalfont)  at low-magnetic fields and high pressures. Red points and green points correspond to the values of T$_M$ at zero field (from Fig.2) and under field above the  critical pressure respectively. Orange dots are the values of  T$_M$ at 7.5 GPa (from Fig.3c), close to the critical pressure, and the black points indicate where ac-susceptibility curves detected no transition down to 1.5 K.}
\end{figure}
Now we discuss why ytterbium systems might favor FM rather than AFM order in contrast to most cerium compounds. One simple idea stems from the recent theoretical works showing how magnetic field can control the valence transition.\cite{Watanabe11} The application of field can induce the first order valence transition (FOVT) accompanied with strong metamagnetic and magnetostriction effects. Conversely if the volume change and valence transition are induced by pressure, the spontaneous magnetization can easily appear. More precisely close to a valence transition FM would be favored by the presence of large nonlocal dynamical deformation of the lattice.\cite{Watanabe11,Harima09} This could explain why ferromagnetism is more common in ytterbium where the valence changes with pressure are much larger than in cerium systems.
We turn now to the consequences of FM rather than AFM order. Ferromagnetism in YbCu$_2$Si$_2$leads us to inquire about the nature of the order of the transition at low temperatures. From the phase diagram it appears already that magnetic order appears rather suddenly at T > 1 K inspiring speculation that the transition could be first-order.\cite{Colombier2009} This is now less surprising as it seems a rather general property of ferromagnets, and so far no example has been found of a FM quantum critical point (QCP), with a second-order transition down to T = 0. The experimental studies of several itinerant ferromagnets like ZrZn$_2$,\cite{Uhlarz04} Co(S$_{(1-x)}$Se$_x$)$_2$,\cite{Goto97} MnSi\cite{Thessieu97} and UGe$_2$ \cite{Huxley00} all show a tricritical point where the paramagnet-ferromagnet transition changes from a second-order to a first-order phase transition when it is driven towards a QCP by applying either external or chemical pressure. These results might suggest that in YbCu$_2$Si$_2$ the transition also becomes first-order near the critical pressure.
 
Another consequence is that FM fluctuations are less favorable than AFM ones for magnetically induced superconductivity.\cite{Monthoux01} This might be one of the reasons that superconductivity is far more elusive in ytterbium systems than in cerium compounds. The first-order transition also changes the picture. In UGe$_2$ which is so far the only system where superconductivity occurs at a FM critical point with a large ordered moment, superconductivity appears only within the FM region. Superconductivity, if it exists in YbCu$_2$Si$_2$, might occur at very low temperature as the 4f  levels in Yb-based compounds are narrower than in Ce-based ones which leads to lower characteristic temperatures, extreme sensitivity to sample quality and an appearance not necessarily where it is expected in the phase diagram.

\section{Conclusions}
Ac susceptibility and ac calorimetry measurements in high-quality single crystals of YbCu$_2$Si$_2$ under high pressure and small magnetic fields strongly suggest that the magnetic order induced for p > 8GPa is FM. Pressure measurements were performed in diamond-anvil cells with argon as pressure-transmitting media and the results confirmed the (\itshape{p,T}\normalfont) phase diagram of YbCu$_2$Si$_2$ previously reported but surprisingly, the magnetic transition temperatures found in this study are significantly higher. We might attribute the difference to the improvement of the crystal quality. 
We have compared the low temperature susceptibility curves with different fits; two Curie laws with $\mu_{eff}$ = 1.25 $\mu_B$ and $\mu_{eff}$ = 4.54 $\mu_B$ respectively largely underestimate the experimental values at low temperature. A Curie-Weiss law with T$_{CW}$ = 2.9 K and $\mu_{eff}$ = 3.52 $\mu_B$ gives the best agreement with the experimental data. Even though this value of the effective magnetic moment is larger than the value expected from Mössbauer spectroscopy, these results point strongly to ferromagnetism. 
Further proof for FM in this system was obtained by superimposing a small magnetic field. The (\itshape p,H,T\normalfont) phase diagram clearly shows that the transition temperature increases with magnetic fields larger than 0.1 T for all pressures we measured and simultaneously the transition is rapidly suppressed under field. These trends are experimentally consistent with the behavior of a FM system under magnetic field. Of course we can not neglect a more complex magnetic structure but a significant FM component is at least observed.
FM seems to appear abruptly at about 8 GPa, suggesting a first-order transition. This is what it has been observed in all ferromagnets up to now: near the vicinity of a QCP the FM transition becomes first order.
Due to the availability now of extremely high quality crystals, YbCu$_2$Si$_2$  is therefore a good candidate to explore in great accuracy the phase diagram around the critical pressure at very low temperature and under magnetic field. It is an experimental challenge due to the rather high pressure involved but could be very rewarding.

\begin{acknowledgments}
We Thank J.-P. Sanchez and J.-L. Tholence for useful discussions. This work was supported by the French National Research Agency (ANR) through the contracts ECCE and DELICE.
\end{acknowledgments}

\bibliography{biblio}

\end{document}